\documentclass[aps,prr,twocolumn,amsmath,amssymb,superscriptaddress,longbibliography]{revtex4-1}
\usepackage[english]{babel}
\usepackage{xcolor}
\usepackage{amssymb}
\usepackage{dcolumn}
\usepackage{bm}
\usepackage{graphicx}
\usepackage{amsmath}
\usepackage{braket}

\usepackage{graphicx}        

\graphicspath{{pict/}{}}

\usepackage[normalem]{ulem}

\usepackage{dcolumn}
\usepackage{bm}
\usepackage[pdfstartview=FitH, CJKbookmarks=true, bookmarksnumbered=true, bookmarksopen=true, colorlinks=true, pdfborder=001, citecolor=blue, linkcolor=blue, urlcolor=blue, linktocpage=true] {hyperref}

\begin{document}

\title{Self-ordered supersolid phase beyond Dicke superradiance in a ring cavity}

\author{Yuangang Deng}
\email{dengyg3@mail.sysu.edu.cn}
\affiliation{Guangdong Provincial Key Laboratory of Quantum Metrology and Sensing $\&$ School of Physics and Astronomy, Sun Yat-Sen University (Zhuhai Campus), Zhuhai 519082, China}

\author{Su Yi}
\email{syi@itp.ac.cn}
\affiliation{CAS Key Laboratory of Theoretical Physics, Institute of Theoretical Physics, Chinese Academy of Sciences, Beijing 100190, China}
\affiliation{CAS Center for Excellence in Topological Quantum Computation, University of Chinese Academy of Sciences, Beijing 100049, China}
\affiliation{School of Physical Sciences, University of Chinese Academy of Sciences, Beijing 100049, China}

\date{\today}

\begin{abstract}
The supersolid phase characterized by the superfluid and long-range spatial periodicity of crystalline order is central to many branches of science ranging from condensed matter physics to ultracold atomic physics. Here we study a self-ordered checkerboard supersolid phase originating from dynamical spin-orbit coupling for a transversely pumped atomic Bose-Einstein condensate trapped in a ring cavity, corresponding to a superradiant anti-Tavis-Cummings phase transition. In particular, an undamped gapless Goldstone mode is observed in contrast to the experimentally realized lattice supersolid with a gapped roton mode for Dicke superradiance. This zero energy mode reveals the rigidity of the self-ordered superradiant phase, which spontaneously breaks a continuous translational symmetry. Our work will highlight the significant opportunities for exploring long-lived supersolid matter by utilizing dynamical spin-orbit coupling in controllable optical cavities.
\end{abstract}

\maketitle
\section{introduction\label{Sec1}}
Supersolidity as a paradigmatic manifestation of novel quantum matter combines two mutually exclusive concepts: the frictionless flow of a superfluid and long-range spatial periodicity of solids~\cite{PhysRevLett.25.1543,RevModPhys.84.759,chan2013overview}. This quantum state spontaneously breaks two continuous $U(1)$ symmetries of the internal gauge symmetry for a superfluid and the translational symmetry for crystal-like periodic density modulation. Despite decades of theoretical  predictions~\cite{PhysRev.106.161,andreev1969quantum,chester1970speculations} and extensive experimental efforts~\cite{kim2004probable,balibar2010enigma,nyeki2017intertwined}, the unambiguously experimental verification of supersolidity in solid helium remains elusive~\cite{PhysRevLett.109.155301}. Ultracold quantum gases with spin-orbit coupling (SOC) ~\cite{NATYJ2011SOCAT, PRLZW2012SOCAT, SCPJW2016SOC, NSHL2016} open the forefront for studying novel coherence effects in many-body physics~\cite{RevModPhys.80.885,RevModPhys.83.863,RevModPhys.83.1523,RevModPhys.91.015005,gross2017quantum},  owing to the versatility of controllable interaction, geometry, and dimensionality. Recently, experimental observations of a stripe phase with supersolidity properties were built by intrinsic interactions in both a Bose-Einstein condensate (BEC) with SOC~\cite{li2017stripe} and axially elongated dipolar quantum droplets~\cite{PhysRevX.9.011051,PhysRevX.9.021012,PhysRevLett.122.130405,guo2019low,tanzi2019supersolid},  which are interesting alternatives of supersolid phases. In these significant advances~\cite{li2017stripe,PhysRevX.9.011051,PhysRevX.9.021012,PhysRevLett.122.130405,guo2019low,tanzi2019supersolid}, the supersolid stripe phase only exists in a narrow parameter regime and reveals weak stripe periodic density modulation.

\begin{figure}[ht]
\centering\includegraphics[width=0.96\columnwidth]{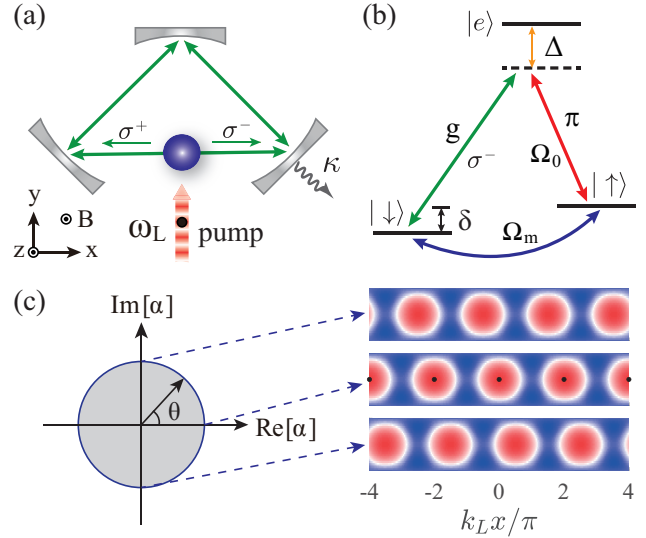}
\protect\caption{(a) Sketch of a BEC for creating dynamical SOC inside a ring cavity. (b) Level diagram. (c) The self-ordered supersolid phase with spontaneously continuous symmetry broken in terms of the cavity amplitude ${\rm Re}[\alpha]$ and ${\rm Im}[\alpha]$ forms a circle. The density profiles for the CB phase are $\lambda$ periodic, corresponding to the positions changing continuously along the $x$ axis by varying phase angle ${\rm{arg}(\alpha)}$. The black dots denote the positions of  $\cos(k_Lx)\cos(k_Ly)=1$.}
\label{scheme}
\end{figure}

Meanwhile, ultracold quantum gases coupled to optical cavities have provided a platform for exploring the periodic order of crystals in controlled environments~\cite{RevModPhys.85.553,Mivehvar:2021lpo,RevModPhys.91.025005}. In particular, an additional collective-emission-induced cooling mechanism could facilitate the experimental investigations of crystal properties in cavity quantum electrodynamics (QED)~\cite{PhysRevLett.89.253003,PhysRevLett.90.063003}. Such cavity-mediated long-range interaction~\cite{Matthew2018,landig2016quantum,PhysRevX.8.011002,Konishi,Muniz20,Wu21,baumann2010dicke,mottl2012roton} and self-ordered dynamical SOC~\cite{PhysRevLett.121.163601,kroeze2019dynamical,PhysRevLett.112.143007,PhysRevA.89.011602,PhysRevA.89.013803,PhysRevA.92.023611} implicate exciting opportunities for creating exotic quantum matter~\cite{PhysRevLett.121.223601,PhysRevLett.120.223602,PhysRevLett.122.193605,PhysRevA.101.063627,PhysRevLett.127.173606,science.aaw4465,Ostermann_2019,PhysRevA.103.023302}. Recently, a self-organized lattice-supersolid phase with discrete ${\cal Z}_2$ symmetry breaking has been observed based on a superradiant Dicke phase transition in experiments~\cite{baumann2010dicke,mottl2012roton,PhysRevLett.121.163601,kroeze2019dynamical}. Furthermore, a supersolid state exhibiting a collective gapless Goldstone excitation has been studied for a spinless BEC coupled to a pair of degenerate modes of the ring cavity~\cite{PhysRevLett.120.123601,PhysRevLett.122.190801,PhysRevLett.124.033601,PhysRevLett.124.143602} or two noninterfering standing-wave optical cavities~\cite{leonard2017supersolid,leonard2017monitoring}. The mechanism for realizing a supersolid phase is that a $U(1)$ symmetry is formed by employing two ${\cal Z}_2$ symmetries with strictly equal couplings for two-mode cavities. An interesting question is \emph{whether there exists an exotic self-ordered supersolid phase induced by dynamical SOC in a single-mode cavity-coupled BEC, corresponding to a crystalline structure by a continuous translational symmetry breaking.} An affirmative answer will not only enrich our knowledge of supersolid phases beyond Dicke superradiance but also provide a nondestructive monitoring tool for self-ordered crystalline orders~\cite{leonard2017supersolid,leonard2017monitoring,PhysRevLett.128.103201}.

In this work, we propose an experimentally accessible scheme to realize a self-ordered supersolid phase by utilizing dynamical SOC in a spin-$1/2$ BEC confined in a ring cavity. Here the dynamical SOC arises from a spatial self-organized ground-state atomic wave function. Above a threshold Raman field, the dynamical SOC leads to a normal superradiant (NSR) phase, a plane-wave (PW) phase, and a checkerboard (CB) supersolid phase existing in a regime with a large range of parameters. In contrast to superradiant Dicke phase transition~\cite{baumann2010dicke,mottl2012roton,PhysRevLett.121.163601,kroeze2019dynamical}, we show that the self-ordered superradiant phases can be fully characterized by the seminal anti-Tavis-Cummings model (TCM). In particular, the supersolid CB phase hosting a gapless Goldstone mode corresponding to spontaneous $U(1)$ symmetry breaking is demonstrated, which is essentially different from the previously realized lattice-supersolid phase with a gapped roton mode~\cite{baumann2010dicke,mottl2012roton,PhysRevLett.121.163601,kroeze2019dynamical}. Compared to pioneer explorations for a scale BEC coupled to a multimode ring cavity~\cite{PhysRevLett.120.123601,PhysRevLett.122.190801,PhysRevLett.124.033601,PhysRevLett.124.143602},  the supersolid phases generated for a cavity-mediated dynamical spin-orbit- coupled spinor BEC with versatile spin degrees of freedom will provide an opportunity to study exotic many-body quantum matter\cite{RevModPhys.80.885,RevModPhys.83.863,RevModPhys.83.1523,RevModPhys.91.015005,gross2017quantum,RevModPhys.85.553,Mivehvar:2021lpo,RevModPhys.91.025005}.

This paper is organized as follows. In Sec.~\ref{Sec2} we introduce our scheme for generating dynamical SOC and derive the Hamiltonian of the system. Section~\ref{Sec3} is devoted to the study of self-ordered quantum phases of cavity-mediated dynamical SOC. Finally, a brief summary is given in Sec.~\ref{Sec4}.

\section{Model and Hamiltonain\label{Sec2}}

We consider an $N_a$ atomic $^{87}$Rb BEC of $F=1$ ground electronic manifold placed inside a high-finesse ring cavity, as shown in Fig.~\ref{scheme}(a). A large bias magnetic field $\mathbf{\mathit{\mathbf{B}}}$ along the $z$ direction is applied to select two hyperfine states labeled as $|\uparrow\rangle=|F=1,m_F=-1\rangle$ and $|\downarrow\rangle=|F=1,m_F=0\rangle$~\cite{NATYJ2011SOCAT}, which also defines the quantization axis. This results in a Zeeman shift $\hbar \omega_Z$ between $|\uparrow\rangle$ and $|\downarrow\rangle$ whose magnetic quantum numbers of the electronic states satisfy $m_{\uparrow}=m_{\downarrow} -1$. To engineer the dynamical SOC, the atomic transition $|\downarrow\rangle\leftrightarrow |e\rangle$ ($|\uparrow\rangle\leftrightarrow |e\rangle$) is coupled to the cavity (transverse pump) field [Fig.~\ref{scheme}(b)], which yields a cavity-mediated two-photon Raman process. Here $\delta$ is the tunable two-photon detuning and $\kappa$ is the decay rate of the ring cavity. Moreover, $|\uparrow\rangle$ and $|\downarrow\rangle$ states are also coupled {\color{blue}to} an external radio frequency (rf) field with coupling strength $\Omega_m$. As we show show below, this spatially independent rf field will play an important role in realizing self-ordered crystalline structures in cooperation with dynamical SOC.

As can be seen, atoms are coherent illuminated by a $\pi$-polarized standing-wave pump field propagating along the $y$ axis. The corresponding Rabi frequency is $\Omega_0(y)=\Omega_0\cos(k_Ly)$, where $k_L= 2\pi/\lambda$ is the wave vector of the laser field with $\lambda$ being the wavelength. Due to the collective Bragg scattering, the optical cavity could  support $\sigma$-polarized and $\pi$-polarized photon fields, simultaneously. Under the conditions $|\delta/\omega_Z|\ll 1$ and $|\omega_Z/\kappa|\gg 1$, the optical cavity only supports a resonant $\sigma$-polarized photon mode with magnetic quantum numbers satisfying $\Delta m=\pm1$, where the $\pi$-polarized photon mode in the far dispersive regime is suppressed~\cite{PhysRevLett.112.143007}. Interestingly, we should point out that the condensate only couples to the $\sigma^-$-polarized photon field with single atom-photon coupling $g(x)=g e^{ik_Lx}$ along the $x$ axis, as shown in Fig.~\ref{scheme}(b). In general, another counterpropagating running wave with a $\sigma^+$-polarized cavity mode along the negative $x$ axis could also exist by the reverse coherent scattering process~\cite{Ostermann_2019,PhysRevLett.120.123601,PhysRevLett.122.190801,PhysRevLett.124.033601,PhysRevLett.124.143602}. However, the Bragg scattering into the $\sigma^+$-polarized cavity mode is decoupled from our laser configurations and can be safely ignored~\cite{PhysRevA.89.011602,bao2012efficient}.

In contrast to the experimentally observed supersolid phase with sufficiently equal pump strengths for the two modes of the ring cavity~\cite{PhysRevLett.124.143602} and two standing-wave cavities~\cite{leonard2017supersolid,leonard2017monitoring}, the advantages of our proposal of utilizing a single ring mode will facilitate the experimental feasibility for engineering quantum states with a simpler laser configuration. More importantly, the mechanism of generating a self-ordered supersolid phase for a spinor condensate is essentially different from the pioneering studies for a spinless BEC coupled to a multimode ring or standing-wave optical cavities~\cite{PhysRevLett.120.123601,PhysRevLett.122.190801,PhysRevLett.124.033601,PhysRevLett.124.143602,leonard2017supersolid,leonard2017monitoring}.

After adiabatically eliminating the atomic excited state $|e\rangle$ in the large atom-pump detuning limit $|\Delta|\gg \{g,\Omega_{0}\}$, the many-body Hamiltonian of the cavity condensate reduces to (see Appendix \ref{sm_model} for more details)
\begin{align}
\hat {\cal H}_0=&\sum_{\sigma\sigma'}\int d{\mathbf r}\hat\psi_{\sigma}^{\dag}({\mathbf r})[\hat h_{\sigma\sigma'}+V_{b}({\mathbf r})\delta_{\sigma\sigma'}]\hat\psi_{\sigma'}({\mathbf r}) + \hbar\Delta_c\hat{a}^{\dag}\hat{a} \nonumber \\
+&\sum_{\sigma \sigma'}g_{\sigma\sigma'}\int d{\bf r}\hat{\psi}^\dag_{\sigma}({\bf r})\hat{\psi}^\dag_{\sigma'}({\bf r})\hat{\psi}_{\sigma'}({\bf r})\hat{\psi}_{\sigma}({\bf r}),\label{manyh}
\end{align}
where $\hat\psi_{\sigma=\uparrow,\downarrow}$ denotes the annihilation bosonic atomic field operator for spin-$\sigma$ atom, $\hat a$ is the annihilation operator of the cavity mode, $\Delta_c$ is the pump-cavity detuning, and $V_{b}({\mathbf r})$ is the trapping potential. The two-body contact interaction $g_{\sigma\sigma'} = 4\pi\hbar^2 a_{\sigma\sigma'}/M$ with $M$ being the mass of the atom and $a_{\sigma\sigma'}$ being the $s$-wave scattering lengths between the intraspecies ($\sigma=\sigma'$) and interspecies ($\sigma\neq\sigma'$) spin atoms. Moreover, the effective single-particle Hamiltonian reads
\begin{eqnarray}
\hat {\boldsymbol h} =\frac{{\mathbf p}^{2}}{2M}\hat I+
\hbar\left(\!
\begin{array}{cc}
\hat{M}_0(y)+ {\delta}/2 & \hat{M}_-(x,y)   \\
\hat{M}_-^\dag(x,y) & \hat{M}_0(y)-{\delta}/2
\end{array}
\!\right)\!,\;\label{singleh}
\end{eqnarray}
where $\hat{M}_0(y)=U_0\hat{a}^{\dag}\hat{a}+ U_y\cos^{2}(k_Ly)$ is the optical lattice with $U_0=-g^2/\Delta$ ($U_y=-\Omega^2_0/\Delta$) being the optical Stark shift of the cavity (pump) field and $\hat{M}_-(x,y)=\Omega_m +\Omega\cos(k_Ly)\hat{a} e^{ik_Lx}$ is the Raman coupling with $\Omega = -g\Omega_0/\Delta$ corresponding to the maximum scattering rate. Compared with the extensively studied superradiant lattice using standing-wave cavity~\cite{baumann2010dicke,mottl2012roton,PhysRevLett.121.163601,kroeze2019dynamical}, we remark that the $\hat{M}_0$ term induced by the running-wave ring cavity respects the continuous translational symmetry along the cavity axis. Moreover, the spin-flip term $\hat{M}_-(x,y)$ contains an interesting dynamical SOC~\cite{PhysRevLett.112.143007,PhysRevA.89.011602} engineered by the interference between the quantized cavity photon and the classical pump field.

In the far dispersive regime, $|\Delta_c|\gg \{\Omega,\Omega_m\}$, the cavity field $\hat{a}$ can be adiabatically eliminated and replaced by a steady-state solution since its dynamical evolution is much faster than the external atomic motion~\cite{RevModPhys.85.553,Mivehvar:2021lpo}. By introducing the parameter ${\Xi}  = \langle
{\psi}_\downarrow|\cos(k_Ly)e^{-ik_Lx}|{\psi}_\uparrow\rangle$, the intracavity amplitude can be expressed as $\alpha=\langle\hat a\rangle= {\Omega{{\Xi}}}/({-\tilde{\Delta}_c +
i\kappa})$ with $\tilde{\Delta}_c =(\Delta_c+U_0N_a)$ being the $N_a$ dependent effective dispersive shift of the cavity (see Appendix \ref{sm_model} for more details). It is clear that a finite $\alpha$ yielding $\Xi\neq0$ characterizes the self-ordered superradiant phase transition. We should note that the quantum noise of the cavity can be ignored since $|\tilde{\Delta}_c/\kappa|\gg 1$.

To proceed further, the cavity-condensate can be fully characterized by a seminal anti-TCM Hamiltonian (see Appendix \ref{appA} for more details)
\begin{align}
\hat {\cal H}/\hbar &= \tilde{\Delta}_c\hat{a}^\dag\hat{a} +
\omega_0\hat{J}_{z}+
\frac{\Omega}{\sqrt{2}}[\hat{a}\hat{J}_{-}  + {\rm H.c.}],\label{JCM}
\end{align}%
where $\hat b_{\downarrow}$ and $\hat b_{\uparrow}$ represent the bosonic mode operators and the populated number of ground state in $|\uparrow\rangle$ is $N_\uparrow = \hat b_{\uparrow}^\dag\hat b_{\uparrow} + \hat b_{\downarrow}^\dag\hat b_{\downarrow}  =N_a(\sqrt{\Omega_m^2+\delta^2/4}-\delta/2)/\sqrt{4\Omega_m^2+\delta^2}$ without superradiance ($\alpha=0$). Here $\omega_0= 2E_L/\hbar-\delta$ is the detuning of the atomic field, and $\hat{J}_{-} = \hat b_{\uparrow}^\dag\hat b_{\downarrow}$ and $\hat{J}_{z} = (\hat b_{\downarrow}^\dag\hat b_{\downarrow} -\hat
b_{\uparrow}^\dag\hat b_{\uparrow})/2$ are the collective spin operators. Taking into account the cavity decay $\kappa$, the analytic critical Raman coupling for superradiance reads (as discussed in Appendix \ref{appA})
\begin{align}
\Omega_{\rm cr} =\sqrt{{2(\tilde{\Delta}_c^2+\kappa^2)^{1/2}\omega_{0}/N_\uparrow}}.\label{PB}
\end{align}

\begin{figure}[ht]
\centering\includegraphics[width=0.96\columnwidth]{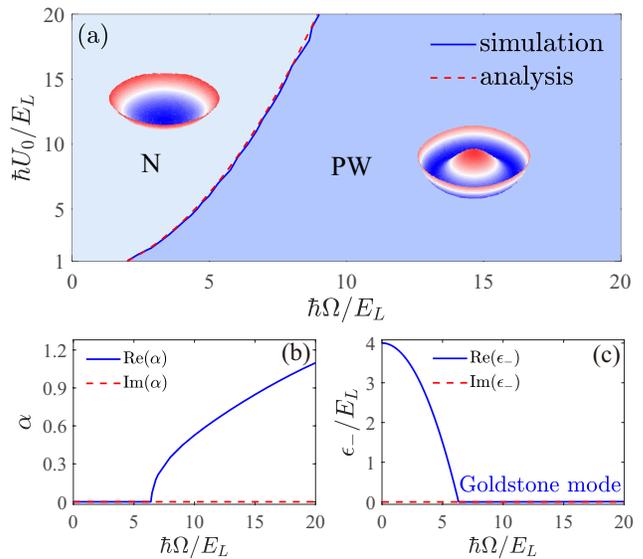}
\protect\caption{(a) Ground-state phase diagram on  $\Omega$-$U_0$ parameter plane with $\Omega_m=0$. The solid (dashed) line denotes the numerical (analytical) result of superradiant phase boundary. The inset shows the effective potential for the N phase with a bowl shape and the PW phase with a Mexican hat. (b) The cavity amplitude $\alpha$ and (c) lower branch of collective excitation $\epsilon_-$ as a function of $\Omega$ with $U_0=10 E_L/\hbar$. The nonzero $\alpha$ and a gapless Goldstone mode correspond to a spontaneously continuous $U(1)$ symmetry broken.}
\label{phase}
\end{figure}

In the single recoil scattering limit~\cite{baumann2010dicke,mottl2012roton,PhysRevLett.121.163601,kroeze2019dynamical}, the microscopic picture of superradiance coherently transfers the atomic motional ground state $|k_{x}=0,k_{y}=0\rangle$ for spin-$\uparrow$ atoms to the excited momentum states $|-k_L,\pm k_L\rangle$ for spin-$\downarrow$ atoms via cavity-emerged dynamical SOC. We should emphasize that the anti-TCM Hamiltonian (\ref{JCM}) is obtained in the absence of two-body collisional atom-atom interaction. In fact, we have examined the threshold for superradiant phase transition and found it is independent of the two-body contact interaction in the single recoil scattering approximation~\cite{baumann2010dicke,mottl2012roton,PhysRevLett.121.163601,kroeze2019dynamical}.

\section{Results\label{Sec3}}

We explore the ground-state structures of the cavity-condensate system with steady-state solution of the cavity photon self-consistently by solving the Gross-Pitaevskii equations using the mean-field theory. To this end, the atomic field operators $\hat{\psi}_{\alpha}$ are replaced by the condensate wave function $\psi_{\alpha}=\langle \hat{\psi}_{\alpha}\rangle$ and the steady-state photon amplitude $\alpha=\langle \hat{a}\rangle$ is self-consistently determined by the condensate wave function. To utilize the
familiar imaginary time evolution, we can obtain the ground states of the condensate wave function by numerically minimizing the free energy functional ${\cal F}[\psi_{\uparrow}, \psi_{\downarrow}] = \langle \hat {\cal H}_0\rangle$. Specifically, we assume an $N_a=10^5$ BEC initially prepared in the $|\uparrow\rangle$ state and confined in a quasi-two-dimensional circular optical box trap~\cite{PhysRevLett.110.200406,Nature2016}, which generates a uniform potential with preserving continuous translational symmetry. For $F=1$ $^{87}$Rb atoms, $s$-wave scattering lengths for the collisional interactions are $a_{\downarrow\downarrow}=a_{\uparrow\downarrow}\approx a_{\uparrow\uparrow}=100.4\,a_B$ with $a_B$ being the Bohr radius~\cite{deng2021universal,PhysRevLett.88.093201}. Indeed, we find that the emerged self-ordered superradiance phases are not dependent on the specific choice the values of $a_{\sigma\sigma'}$. The singe-photon recoil energy is $E_L/\hbar = 3.53 ~{\rm kHz}(2\pi)$ corresponding to $\lambda=803.2$ nm with respect to the wavelength of the atomic transition. In numerical simulations, we adopt the cavity decay rate $\kappa=100 E_L/\hbar$, two-photon detuning $\delta=-2E_L/\hbar$, and pump-cavity detuning $\Delta_c=-U_0N_a/2$. Then the independent controllable parameters are Stark shift $U_0$, Rabi frequency $\Omega$, and rf field $\Omega_m$.

Figure \ref{phase}(a) shows the phase diagram of the condensate with cavity-emerged dynamical SOC on the $\Omega$-$U_0$ parameter plane without rf field. A superfluid phase with zero intracavity amplitude is denoted by ``N''. The quantum phase transition from N to self-ordered PW phase can be ascribed to cavity superradiance. The $N_a$-independent analytic threshold for superradiance satisfies $\Omega_{\rm cr} = \sqrt{U_0\omega_0}$, ignoring the cavity decay and atom collision (dashed line), which is in high agreement with numerical result (solid line). We check that the numerical results for generating self-ordered superradiant phases are robust against the variations of the $s$-wave scattering lengths $a_{\sigma\sigma'}$, as the cavity-mediated long-range spin-exchange interaction for the atomic field is much larger than the two-body contact interaction. The crucial feature of superradiance is that the cavity amplitude corresponds to a spontaneous symmetry breaking from vacuum ($\alpha=0$) to a finite value ($\alpha\neq 0$), as shown in Fig.~\ref{phase}(b).

In contrast to Dicke superradiance with ${\cal Z}_2$ symmetry $\hat{a}\rightarrow -\hat{a}$ and $\hat{J}_{\pm}\rightarrow -\hat{J}_{\pm}$ ~\cite{baumann2010dicke,mottl2012roton,PhysRevLett.121.163601,kroeze2019dynamical}, the anti-TCM of Eq.~(\ref{JCM}) possesses a continuous $U(1)$ symmetry characterized by the operator ${\cal R}_{\theta} =\exp[i\theta(\hat{a}^\dag\hat{a}-\hat{J}_z)]$, which yields ${\cal R}_{\theta}^\dag(\hat{a},\hat{J}_-
,\hat{J}_+){\cal R}_{\theta} =(\hat{a} e^{-i\theta},\hat{J}_-e^{i\theta},\hat{J}_+e^{-i\theta})
$~\cite{PhysRevE.67.066203,PhysRevLett.112.173601}. This symmetry will be spontaneously broken with self-ordered anti-TCM superradiance. We demonstrated that the effective potential will change from a minimum in the origin to a shape of a Mexican hat with a circular valley of degenerate minima when the superradiant phase transition occurs (see Appendix \ref{appB} for more details). Remarkably, a gapless Goldstone mode of the low-energy excitation is confirmed [Fig.~\ref{phase}(c)], in contrast to the gapped roton mode for Dicke superradiance~\cite{mottl2012roton}. Interestingly, this zero energy mode is roughly undamped even with nonzero cavity dissipation since $\kappa/\tilde{\Delta}_c\sim 10^{-3}\ll 1$ in our simulation. More discussions are found in Appendix \ref{appC}.

\begin{figure}[ht]
\centering\includegraphics[width=0.96\columnwidth]{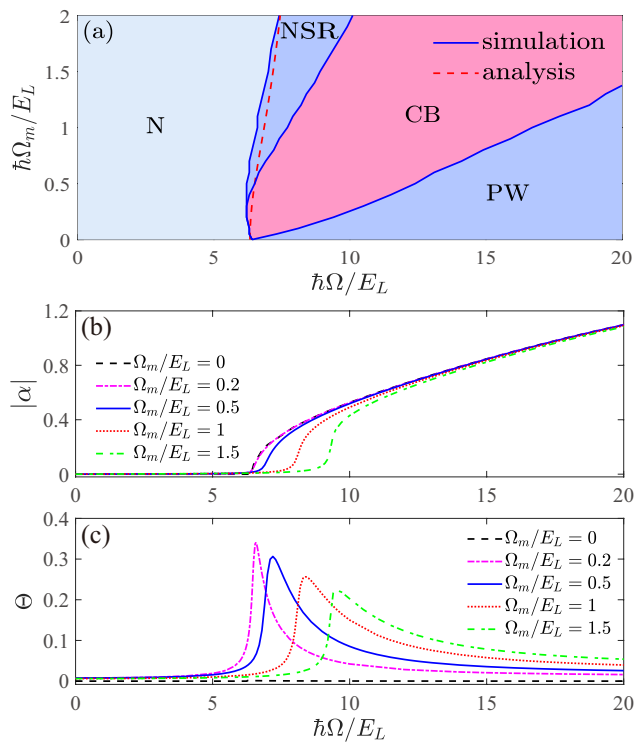}
\protect\caption{(a) Emergence of self-ordered superradiant phases on the $\Omega$-$\Omega_m$ parameter plane with $U_0=10E_L/\hbar$.  The $\Omega$ dependence of order parameters (b) $|\alpha|$ and (c) $\Theta$ for different values of $\Omega_m$. }
\label{orders}
\end{figure}

\begin{table}[!htbp]
{\color{black}\caption{The order parameters for the self-ordered superradiant phases. Here, ``homogeneous'' and ``$\lambda$-periodic'' denotes the relative phase of condensate wave function $\Delta \phi$ exhibiting a structureless phase profile and staggered $\lambda$-periodic phase modulation along the $x$ axis, respectively.}
\label{ords}
\begin{tabular}{|l|l|l|l|}
\hline
\parbox[c]{1.2cm}{Phases}  & \parbox[c]{2.2cm}{NSR} & \parbox[c]{2.2cm}{CB}  & \parbox[c]{2.2cm}{PW} \\
\hline
\parbox[c]{1.2cm}{$|\alpha|$}  &  \parbox[c]{2.2cm}{$>0$} & \parbox[c]{2.2cm}{$>0$} & \parbox[c]{2.2cm}{$>0$}
\\
\hline
\parbox[c]{1.2cm}{$\Theta$}  &  \parbox[c]{2.2cm}{$<0.05$} & \parbox[c]{2.2cm}{$>0.05$} & \parbox[c]{2.2cm}{$<0.05$}
\\
\hline
\parbox[c]{1.2cm}{$\Delta \phi$}  &  \parbox[c]{2.2cm}{homogeneous} & \parbox[c]{2.2cm}{$\lambda$ periodic} & \parbox[c]{2.2cm}{$\lambda$ periodic}
\\
\hline
\end{tabular}}
\end{table}

Figure \ref{orders}(a) summarizes the quantum phases on the $\Omega$-$\Omega_m$ parameter plane with fixing $U_0=10 E_L$. As can be seen, an additional two self-ordered superradiant phases in the presence of $\Omega_m$ are observed. To further characterize these phases, we introduce an order parameter defined by ${\Theta}  = \langle
{\psi}_\downarrow|\cos(k_Lx)\cos(k_Ly)|{\psi}_\downarrow\rangle/N_{\downarrow}$, which measures the configurations of periodic density modulation. Then three self-ordered superradiant phases are immediately distinguished. Explicitly, the order parameters for superradiant phases including NSR, CB, and PW are given in Table \ref{ords}. The cavity amplitude can be well used to characterize the phase boundary between the N phase ($\alpha=0$) and the self-ordered superradiance phases ($\alpha\neq0$). An can be seen, the CB phase hosts a macroscopic value of order parameter ${\Theta}$ ($>0.05$), indicating the emergence of a strong self-ordered crystalline structure, which corresponds with the photon amplitude roughly satisfying $|\alpha|>0.05$ as well. Both the NSR and PW phases with $\Theta<0.05$ indicate a weak stripe periodic density modulation along the cavity axis. We should emphasize that the critical value of $\Theta=0.05$ is considered which is limited by the numerical simulations for a finite size of condensate. Indeed, the order parameter should satisfy $\Theta=0$ for phase transition between CB and NSR (PW) in the thermodynamic limit. We remark that the phase diagram in Fig.~\ref{orders}(a) remains qualitatively unchanged against the variations of the critical value of $\Theta$.

Furthermore, we find that the NSR phase possesses a small cavity amplitude in contrast to the PW phase with a large value of $|\alpha|$. Therefore, the PW phase possesses a periodic phase modulation along the $x$ axis induced by the dynamical SOC compared to the NSR phase with a structureless phase profile. For a large $\Omega_m$, a NSR phase exists around superradiant the phase boundary instead of the PW phase with $\Omega_m=0$. For a small $\Omega_m$, the PW phase is dominated by cavity-mediated dynamical SOC at a large Raman coupling. As for the moderate $\Omega$ and $\Omega_m$, an interesting CB phase exhibiting a spatial periodicity of crystalline order for density profiles emerges, which demonstrates a self-ordered supersolid phase in combination with the gapless Goldstone mode [Fig.~\ref{phase}(c)].

Remarkably, the threshold of superradiant scattering is immune to small values of $\Omega_m$, since the rf field is not directly coupled to the cavity mode in the single recoil scattering limit~\cite{baumann2010dicke,mottl2012roton,PhysRevLett.121.163601,kroeze2019dynamical}.
For a further increasing $\Omega_m$, the numerical results show that the phase boundary of the superradiance ($\Omega_{\rm cr} \sim \sqrt{1/N_\uparrow}$) shifts along the direction of increasing $\Omega$. The reason is that $N_{\uparrow}$ gradually decreases with {\color{blue} an} increasing $\Omega_m$, corresponding to the enhancement of the mixing between two internal spin components (see Appendix \ref{appA}). This result can be well understood from the analytical solution of Eq.~(\ref{PB}) originating from anti-TCM superradiance [dashed line in Fig.~\ref{orders}(a)]. As for the photon amplitude, $|\alpha|$ obviously  shifts along the direction of $\Omega$ for the NSR phase at large $\Omega_m$, as displayed in Fig.~\ref{orders}(b). However, we should emphasize that $|\alpha|$ is insensitive to the varying of $\Omega_m$ when $\Omega$ is far away from the critical value of $\Omega_{\rm cr}$. These results reveal that the self-ordered superradiant phases could be well characterized by the anti-TCM even for nonzero weak $\Omega_m$. But for a strong rf field, e.g., $\hbar\Omega_m/E_L\gg 1$, the system evolution will deviate from the anti-TCM with respect to the single recoil scattering approximation being invalid~\cite{baumann2010dicke,mottl2012roton,PhysRevLett.121.163601,kroeze2019dynamical}.

In Fig.~\ref{orders}(c), we plot the $\Omega$ dependence of ${\Theta}$ for different values of $\Omega_m$. It is clear that the CB phase corresponds to a large value of $\Theta$, which indicates a strong periodic density modulation. Especially for a weak rf field, $\Theta$ exhibits an essentially different lineshape in contrast to the result of the straight line for $\Omega_m=0$, albeit $\alpha$ shows exactly the same behavior. As $\Omega_m$ is increasing, the peak values of ${\Theta}$ decreases gradually with its position shifting along the direction of $\Omega$. Furthermore, the value of ${\Theta}$ for both NSR and PW phases are very small, which demonstrates the structureless density profiles for the condensate wave function. We also check that the NSR phase exhibits a homogeneous phase patten for the condensate wave function in contrast to the PW phase (see Appendix \ref{appA} for more details).

\begin{figure}[ht]
\centering\includegraphics[width=0.99\columnwidth]{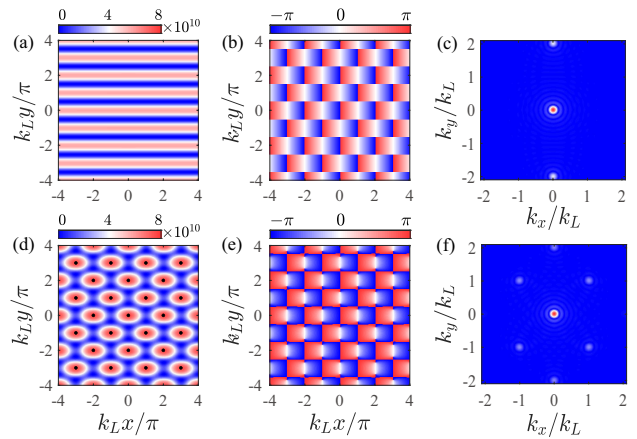}
\protect\caption{The condensate wave functions for the self-ordered PW phase (upper row) with $(U_0,\Omega,\Omega_m)=(10,7.5,0)E_L/\hbar$ and the CB phase (lower row) with $(U_0,\Omega,\Omega_m)=(10,7.5,0.5)E_L/\hbar$. Columns 1 and 2 denotes density (units of cm$^{-2}$) $\rho_{\downarrow}$ and relative phase $\Delta \phi$, respectively. The black dots in (d) denote the positions of sites satisfying $\cos(k_Lx)\cos(k_Ly) = 1$. Column 3 shows the momentum distribution of the $|\downarrow\rangle$ state with the peak areas representing high atomic density. }
\label{density}
\end{figure}

To gain more physical insight, we plot the real-space and momentum-space distributions of the ground states for PW and CB phases in Fig.~\ref{density}. In the parameter regimes of our numerical simulation, the $|\uparrow\rangle$ state is always dominantly populated. Therefore the self-organized structure of condensate wave functions appears in the less populated $|\downarrow\rangle$ component for minimizing kinetic energy. For the PW phase, the density $\rho_{\downarrow}=|{\psi}_{\downarrow}|^2$ with $\psi_\sigma=\langle\hat{\psi}_\sigma\rangle$ exhibits a $\lambda/2$-periodic stripe when applying the $y$-direction optical lattice of the pump field. The relative phase of the condensate wave function satisfies $\Delta \phi=\arg(\psi_\uparrow)-\arg(\psi_\downarrow)= -k_Lx$ ($\pi-k_Lx$) when $\cos(k_Ly)=1$ ($-1$). The staggered $\lambda$-periodic phase modulation that appears along the $x$ axis is determined by the effective transverse magnetic field with ${\bf B}_{\perp}\propto [{\rm{Re}}(\hat{M}_-),-{\rm{Im}}(\hat{M}_-)]$ for a finite $\alpha$~\cite{PhysRevLett.112.143007}. As for the momentum-space distribution, we find the momentum peaks at $|0,\pm 2 k_L\rangle$ along the $y$ axis in addition to a zero momentum for the PW phase, as shown in Fig.~\ref{density}(c).

Figures \ref{density}(d)-\ref{density}(f) show the typical results the of supersolid the CB phase. It is clear that the density profile of CB phase exhibits a spatial periodicity of crystalline order [Fig.~\ref{density}(d)]. The peak density is located at the positions satisfying $\cos(k_Lx)\cos(k_Ly) = 1$ when ${\rm{Re}(\alpha)}>0$ and ${\rm{Im}(\alpha)}=0$. The density positions of the CB phase depend on the atomic wave function via the order parameter $\alpha$, which is directly connected to the phase angle $\rm{arg}(\alpha)$ in the ${\rm Re}[\alpha]$-${\rm Im}[\alpha]$ plane [Fig.~\ref{scheme}(c)]. Remarkably, the positions change continuously along the cavity axis, which corresponds to the anti-TCM superradiance with spontaneous $U(1)$ symmetry breaking. Moreover, $\Delta \phi$ for the CB phase exhibits a similar $\lambda$-periodic phase modulation compared to the PW phase [Fig.~\ref{density}(e)]. Due to the interference between the zero momentum $|0,0\rangle$ of the rf field and $|-k_L,\pm k_L\rangle$ momentum generated by dynamical SOC, the atomic momentum distribution is observed to undergo a striking change with adding the momentum components at $|\pm k_L,\pm k_L\rangle$ [Fig.~\ref{density}(f)]. This result is in agreement with the checkerboard periodic modulation for the density profile associated with the macroscopic value of the order parameter ${\Theta}$, which also demonstrates the validity of the single recoil scattering approximation~\cite{baumann2010dicke,mottl2012roton,PhysRevLett.121.163601,kroeze2019dynamical}. Finally, the supersolid properties for the self-ordered CB phase are unambiguously demonstrated with a combination of spatial periodicity of crystalline order and gapless Goldstone mode [Fig.~\ref{phase}(c)] in the superfluid quantum gases.

We remark that the mechanism for constructing a self-ordered crystalline structure is different from the experimentally realized self-organized lattice supersolid for Dicke superradiance~\cite{baumann2010dicke,mottl2012roton,PhysRevLett.121.163601,kroeze2019dynamical}. As for experimental feasibility, the self-ordered superradiant phases between PW and CB states can be distinguished by measuring the atomic momentum distribution via spin-sensitive absorption images~\cite{PhysRevLett.121.163601}. The NSR and PW phases can be readily distinguished by measuring the cavity amplitude or phase structure of the condensate wave function. In addition, the robustness of the Goldstone mode immune to the cavity dissipation can be detected using spectroscopic measurement~\cite{leonard2017monitoring}. The advantages of our proposal will facilitate monitoring the quantum phase transitions by utilizing the inherent leakage of the cavity~\cite{leonard2017supersolid,leonard2017monitoring,PhysRevLett.128.103201}. This is in contrast to the traditional method where the density modulation is detected using Bragg scattering~\cite{li2017stripe}. Finally, we should emphasize that the supersolid CB phase is absent when $\Omega_m=0$. This can be attributed to the threshold of superradiant $\Omega_{\rm cr}$ ($>2E_L$) being much larger than the existing density stripe regime with $\Omega<0.2 E_L$ for $F=1$ $^{87}$Rb atoms~\cite{PhysRevLett.121.163601,kroeze2019dynamical,PhysRevLett.112.143007}.

\section{Conclusion\label{Sec4}}

Based on state-of-the-art cavity QED, we propose an experimental scheme corresponding to a simpler laser configuration to realize a self-ordered supersolid phase in a BEC trapped within a ring cavity. The superradiance arises from the cavity-emerged dynamical SOC with a continuous translational symmetry breaking associated with the anti-TCM phase transition. In particular, it was shown that the observed supersolid CB phase with a large periodic density modulation possesses an undamped gapless Goldstone mode even in the presence of the cavity dissipation, which is in contrast to the experimentally observed superradiance of a Dicke lattice supersolid with a gapped excitation~\cite{baumann2010dicke,mottl2012roton,PhysRevLett.121.163601,kroeze2019dynamical}. Remarkably, the mechanism of our proposal that uses spin degrees of freedom for a spinor BEC is essentially different from the recently significant advances for scale BEC coupled to multimode optical cavities~\cite{PhysRevLett.120.123601,PhysRevLett.122.190801,PhysRevLett.124.033601,PhysRevLett.124.143602,leonard2017supersolid,leonard2017monitoring}. Compared to the experimentally realized supersolid phase with stripe periodic density modulation in a BEC with SOC~\cite{li2017stripe} and dipolar quantum gases with intrinsic long-range interaction~\cite{PhysRevX.9.011051,PhysRevX.9.021012,PhysRevLett.122.130405,guo2019low,tanzi2019supersolid}, our scheme of employing cavity-induced dynamical SOC could provide a versatile platform for creating and nondestructively detecting intriguing quantum phases in cavity QED and introduce new capabilities into quantum simulations~\cite{RevModPhys.85.553,Mivehvar:2021lpo,RevModPhys.91.025005}. Interestingly, the controllable strongly cavity-mediated long-range spin-exchange interaction for atomic fields can be derived by integrating out the cavity field (see Appendix \ref{sm_model} for more details), which could facilitate the study of complex phenomena in quantum matters~\cite{PhysRevLett.121.223601,PhysRevLett.120.223602,PhysRevLett.122.193605,PhysRevA.101.063627,PhysRevLett.127.173606,science.aaw4465} and high-precision quantum sensing and metrology~\cite{PhysRevLett.122.190801,PhysRevLett.113.073003,PhysRevLett.124.193602}.

\begin{acknowledgments}
{This work was supported by the National Key R$\&$D Program of China (Grants No. 2018YFA0307500 and  No. 2017YFA0304501), NSFC (Grants No. 11874433, No. 12274473, No. 12135018, No. 11674334, No. 11974363, and No. 11947302), and the Key-Area Research and Development Program of GuangDong Province under Grant No. 2019B030330001.}
\end{acknowledgments}

\appendix

\section{Cavity-condensate Hamiltonian\label{sm_model}}

In this appendix we present the details on the derivation of the effective cavity-condensate Hamiltonian for the displayed laser configurations and level diagram in Figs.~\ref{scheme}(a) and \ref{scheme}(b) of the main text.
First, ignoring the two-body collisional atom-atom interaction, the single-particle Hamiltonian of the atom-cavity system under the rotating-wave approximation can be written as
\begin{eqnarray}
\hat {\boldsymbol h}_0/\hbar &= & \Delta_{c}\hat{a}^\dag\hat{a} + \delta
\hat{c}_\uparrow^\dag\hat{c}_\uparrow + \Delta
\hat{e}^\dag\hat{e}\nonumber \\
&+& \bigg(\Omega^*_0({
y})\hat{c}_{\uparrow}^{\dag}\hat{e} + g^*({
x})\hat{a}^\dag\hat{c}_{\downarrow}^\dag \hat{e} + \Omega_m\hat{c}_{\downarrow}^\dag\hat{c}_\uparrow \bigg)+ {\rm H.c.}, \nonumber \\
\end{eqnarray}
where $\hat a$ is the annihilation operator of the cavity field, and ${\hat c}_{\sigma}$ (${\hat e}$) is the annihilation operator of the atomic field for the ground (excited) state. Here $\Omega_0(y)= \Omega_0\cos(k_Ly)$ is the Rabi frequency, $g(x)=ge^{ik_Lx}$ is the single atom-photon coupling, $\Omega_m$ is a spatially independent Rabi coupling for the radio frequency (rf) field which is equivalent to the effective in-plane magnetic field, $\Delta_c$ is the pump-cavity detuning, $\delta$ is the two-photon detuning, and $\Delta$ is the tunable atom-pump detuning.

Taking into account the atomic spontaneous emissions and cavity decay, the Heisenberg equations of motion for cavity and atomic field operators are given by
\begin{eqnarray}
i\dot{\hat c}_{\uparrow} &=& \Omega^*_0(y)\hat{e} + \delta{\hat c}_{\uparrow} + \Omega_m^* {\hat c}_{\downarrow} , \nonumber \\
i\dot{\hat c}_{\downarrow} &=& g^*(x)\hat{a}^\dag\hat{e}  + \Omega_m {\hat c}_{\uparrow} , \nonumber \\
i\dot{\hat e} &=& (\Delta-i\gamma)\hat{e}+ \Omega_0(y){\hat c}_{\uparrow}+g(x)\hat{a}{\hat c}_{\downarrow} , \nonumber \\
i\dot{\hat a} &=& (\Delta_c -i\kappa) \hat{a} + g^*(x)\hat{c}_\downarrow^\dag
\hat{e}, \label{evolution}
\end{eqnarray}
where $\kappa$ is the cavity decay rate and $\gamma$ is the atomic spontaneous emission rate for excited state.

In the large atom-pump detuning limit, e.g., $|\Delta|\gg \{g,\Omega_0,\kappa,\gamma\}$, we can adiabatically eliminate the electronically excited state by setting $i\dot{\hat{e}} =0$, which yields
\begin{align} \label{eli}
\hat{e} &\approx -\frac{1}{\Delta}\bigg(\Omega_0(y){\hat c}_{\uparrow}+g(x)\hat{a}{\hat c}_{\downarrow}\bigg),
\end{align}%
with ignoring the spontaneous decay of the electronic excited state with $|\gamma/\Delta|\ll 1$. Inserting the expression of Eq.~(\ref{eli}) into evolution equation (\ref{evolution}), then the single-particle Hamiltonian for a pseudospin-$1/2$ atom reads (all over
$\delta/2$ energy offset)
\begin{widetext}
\begin{eqnarray}
\hat {\boldsymbol h} =\frac{{\mathbf p}^{2}}{2M}\hat I +
\hbar\left(\!
\begin{array}{cc}
{U_0}\hat{a}^{\dag}\hat{a}  +U_y\cos^{2}(k_Ly) +\delta/2& \Omega_m+\Omega\cos(k_Ly)\hat{a} e^{ik_Lx}   \\
\Omega_m+\Omega\cos(k_Ly)\hat{a}^\dag e^{-ik_Lx} & {U_0}\hat{a}^{\dag}\hat{a}  +U_y\cos^{2}(k_Ly) -\delta/2
\end{array}
\!\right),\!\;
\label{smsingle}
\end{eqnarray}
\end{widetext}
where $M$ is the mass of the atom, $U_0=-g^2/\Delta$ ($U_y=-\Omega^2_0/\Delta$) is the optical Stark shift of the cavity (pump) field. Interestingly, the spin-flip term $\Omega\cos(k_Ly)\hat{a} e^{ik_Lx}$ in $\hat {\boldsymbol h}$ represents a dynamical SOC~\cite{PhysRevLett.112.143007,PhysRevA.89.011602} engineered by the interference between the quantized cavity photon and classical pump field, where $\Omega = -g\Omega_0/\Delta$ is the Raman coupling strength corresponding to the maximum scattering rate. As can be seen, this single-particle Hamiltonian is the same as Eq.~(\ref{singleh}) in the main text.

Then the many-body interaction Hamiltonian for cavity-condensate system is given by
\begin{align}
{\cal \hat{H}}_0 &= {\cal \hat{H}}_c + {\cal \hat{H}}_{a} + {\cal \hat{H}}_{\rm ac},  \label{smmany}
\end{align}%
Explicitly, we have
\begin{widetext}
\begin{align}
{\cal \hat{H}}_c &= \hbar\Delta_c \hat{a}^{\dag}\hat{a} \\
{\cal \hat{H}}_{\rm a} &= \int d {\mathbf r} \{ [U_0\hat{a}^\dag \hat{a}
+ U_y\cos^2(k_Ly) + V_{b}({\mathbf r})]\sum_{\sigma}\hat{\psi}_\sigma^\dag({\mathbf r})\hat{\psi}_\sigma ({\mathbf r}) +\frac{\delta}{2}[\hat{\psi}_\uparrow^\dag({\mathbf r})\hat{\psi}_\uparrow ({\mathbf r}) - \hat{\psi}_\downarrow^\dag({\mathbf r})\hat{\psi}_\downarrow({\mathbf r}) ]\} \nonumber \\
&+\sum_{\sigma \sigma'}g_{\sigma\sigma'}\int d{\bf r}\hat{\psi}^\dag_{\sigma}({\bf r})\hat{\psi}^\dag_{\sigma'}({\bf r})\hat{\psi}_{\sigma'}({\bf r})\hat{\psi}_{\sigma}({\bf r}), \\
{\cal \hat{H}}_{\rm ac} &= \int d {\mathbf r} \{ \Omega\cos(k_Ly)[\hat{a}e^{ik_Lx}\hat{\psi}_\uparrow^\dag({\mathbf r})\hat{\psi}_\downarrow({\mathbf r}) + \hat{a}^\dag e^{-ik_Lx}\hat{\psi}_\downarrow^\dag({\mathbf r})\hat{\psi}_\uparrow({\mathbf r}) ]+\Omega_m[\hat{\psi}_\uparrow^\dag({\mathbf r})\hat{\psi}_\downarrow({\mathbf r}) + \hat{\psi}_\downarrow^\dag({\mathbf r})\hat{\psi}_\uparrow({\mathbf r}) ]\},
\end{align}%
\end{widetext}
where $\hat{\psi}_{\sigma}({\mathbf r})$ denotes the annihilation
bosonic operator for the atomic field. ${\cal \hat{H}}_c$ represents the Hamiltonian of the ring cavity with nonzero cavity dissipation. The term of ${\cal \hat{H}}_a$ denotes the atomic Hamiltonian for a spin-$1/2$ system, which incudes the external trapping potential, effective Zeeman field, and two-body collisional interaction. The last term of ${\cal \hat{H}}_{\rm ac}$ denotes the cavity-condensate interaction originating from the photon superradiance scattering combination of classical pump and quantized cavity fields.

To proceed further, the dynamical equations for the atom and cavity operators take the form
\begin{widetext}
\begin{align}
i\dot{\hat{a}} & = (\tilde{\Delta}_c -
i\kappa)\hat{a} + \Omega{\hat{\Xi}},\nonumber \\
i\dot{\hat{\psi}}_\uparrow & = [U_0\hat{a}^\dag \hat{a} +U_y\cos^2(k_Ly) + V_{b}({\mathbf r}) + \frac{\delta}{2}]\hat{\psi}_\uparrow +[\Omega\cos(k_Ly)e^{ik_Lx}\hat{a} + \Omega_m]\hat{\psi}_\downarrow + (g_{\uparrow\uparrow}\hat{\psi}_\uparrow^\dag \hat{\psi}_\uparrow + g_{\uparrow\downarrow}\hat{\psi}_\downarrow^\dag \hat{\psi}_\downarrow )\hat{\psi}_\uparrow, \nonumber \\
i\dot{\hat{\psi}}_\downarrow & = [U_0\hat{a}^\dag\hat{a} +U_y\cos^2(k_Ly) + V_{b}({\mathbf r}) - \frac{\delta}{2}]\hat{\psi}_\downarrow +[\Omega\cos(k_Ly)e^{-ik_Lx}\hat{a}^\dag+\Omega_m]\hat{\psi}_\uparrow + (g_{\downarrow\downarrow}\hat{\psi}_\downarrow^\dag \hat{\psi}_\downarrow + g_{\uparrow\downarrow}\hat{\psi}_\uparrow^\dag \hat{\psi}_\uparrow )\hat{\psi}_\downarrow,
 \label{smdyn}
\end{align}%
\end{widetext}
where $\tilde{\Delta}_c =(\Delta_c+U_0N_a)$ is the effective cavity detuning and ${\hat{\Xi}}$ is the introduced parameter defined as
\begin{align}
{\hat{\Xi}}  &= \int d {\mathbf r} \cos(k_Ly)e^{-ik_Lx}
\hat{\psi}_\downarrow^\dag({\mathbf r})\hat{\psi}_\uparrow({\mathbf r}), \nonumber
\end{align}%
which characterizes the spatial distribution of atomic fields.

In order to gain some physical insight, it is appropriate to adiabatically eliminate the cavity field in the far dispersive regime with $|\tilde{\Delta}_c/\kappa|\gg 1$. The cavity field quickly reaching a steady state is much faster than the external atomic motion~\cite{baumann2010dicke,mottl2012roton}. The steady-state equation of motion for the cavity field can be formally solved, yielding
\begin{align}
\hat a= \frac{\Omega{\hat{\Xi}}}{-\tilde{\Delta}_c +
i\kappa},
\end{align}%
where $\alpha=\langle\hat{a}\rangle$ is the amplitude of the intracavity photon. Inserting the steady-sate solution $\hat a$ back into the atomic dynamic equation of  Eq.~(\ref{smdyn}), one can derive the cavity-mediated long-range interaction for the atomic operators by integrating out the cavity field:
\begin{widetext}
\begin{align}
\hat{\cal {H}}_{\rm {eff}} = V_I\int d {\bf r} d {\bf r'}
\hat{\psi}_\uparrow^\dag ({\mathbf r})\hat{\psi}_\downarrow{^\dag}({\mathbf r'})\cos(k_Ly)\cos(k_Ly')e^{ik_L(x-x')}
\hat{\psi}_\uparrow ({\mathbf r'})\hat{\psi}_\downarrow({\mathbf r}), \label{many}
\end{align}
\end{widetext}
which represents the long-range spin-exchange interaction with conserving the atomic number in the individual spin state. Here $V_I=-\frac{\tilde{\Delta}_c\Omega^2}{\tilde{\Delta}_c^2 +
\kappa^2}$ ($V_I<0$) is the tunable strength of the cavity-meditated atom-atom interaction. Hamiltonian (\ref{many}) will dominate the density correlations in the atomic condensate with spatial periodicity of $\lambda$ along the pump and cavity directions~\cite{mottl2012roton}. We remark that the controllable cavity-mediated long-range spin-exchange interaction for the atomic field is much larger than the two-body contact interaction. Thus, the threshold of superradiance is robust against the variations of the $s$-wave scattering lengths $a_{\sigma\sigma'}$ in our numerical simulations.

\section{Superradiant phase transition with dynamical SOC \label{appA}}

In this section, we outline the derivation of quantum phase transition for the anti-Tavis Cummings model, in which the quantum phase transition is relevant to superradiance of the optical cavity. In order to calculate the critical thresholds for the superradiance scattering, we just consider the atomic momentum modes interacting significantly scattered by the process of the cavity-induced Raman coupling and neglect the optical Stark shift $U_y\cos^{2}(k_Ly)$ induced by the pump and weak rf field $\Omega_m$. In the single recoil scattering limit~\cite{baumann2010dicke,mottl2012roton,PhysRevLett.121.163601,kroeze2019dynamical}, the other higher Fourier order modes induced by the Raman process of cavity two-body contact atom-atom interaction can be safely neglected. Thus the atomic field operator $\hat\psi_{\sigma}$ ($\sigma=\uparrow, \downarrow$) from a homogeneous BEC initially prepared in the $|\uparrow\rangle$ state can be expanded as
\begin{align}
\hat\psi_{\uparrow}=\sqrt{\frac{1}{{{V}}}}\hat b_{\uparrow}~ {\rm and}~
\hat\psi_{\downarrow}=\sqrt{\frac{2}{V}}\cos(k_Ly)e^{-ik_Lx}\hat b_{\downarrow}, \label{SM_modes}
\end{align}
where $\hat b_{\sigma}$ is bosonic mode operator with $V$ being the volume of condensates and the total atom number is $N_{\uparrow}=\hat b_{\uparrow}^\dag\hat b_{\uparrow} + \hat b_{\downarrow}^\dag\hat b_{\downarrow}$. It is clear that the relevant atomic momentum modes are $|k_{x},k_{y}\rangle=|-
\hbar k,\pm \hbar k\rangle$ for spin-$\downarrow$ atoms and
$|k_{x},k_{y}\rangle=|0,0\rangle$ for spin-$\uparrow$ atoms, respectively.

After inserting the expansion Eq.~(\ref{SM_modes}) into the many-body Hamiltonian (\ref{singleh}), excluding the external trapping potential and two-body  $s$-wave collisional interaction, the emerged anti-TCM is given by
\begin{align}
\hat {\cal H}/\hbar &= \tilde{\Delta}_c\hat{a}^\dag\hat{a} +
\omega_0\hat{J}_{z}+
\frac{\Omega}{\sqrt{2}}[\hat{a}\hat{J}_{-}  + {\rm H.c.}],\label{smantiJC}
\end{align}%
which describes the $N_{\uparrow}$ two-level bosonic atom system coupled to a single-mode optical cavity. Here $\hat b_{\downarrow}$ and $\hat b_{\uparrow}$ represent the bosonic operator corresponding to the relevant atomic momentum modes $|k_{x},k_{y}\rangle=|-k_L,\pm k_L\rangle$ ($|0,0\rangle$) for spin-$\downarrow$ (spin-$\uparrow$) atoms; $\hat{J}_{-} = \hat b_{\uparrow}^\dag\hat b_{\downarrow}$ and $\hat{J}_{z} = (\hat
b_{\downarrow}^\dag\hat b_{\downarrow} - \hat b_{\uparrow}^\dag\hat b_{\uparrow})/2$ are the collective spin operators. $\tilde{\Delta}_c =(\Delta_c+U_0N_a)$ is the effective cavity detuning and $\omega_0= 2E_L/\hbar-\delta$ is the detuning of the atomic field.

Note that the Hamiltonian terms proportional to $\hat{a}\hat{J}_{-}$ (non-rotating-wave coupling terms) do not conserve the number of total excitations. However, since this describe the simultaneous creation or destruction of two excitations, the parity of the excitation number is conserved. Importantly, the non-rotating-wave Hamiltonian (\ref{smantiJC}) possesses a $U(1)$ symmetry characterized by the action of the operator
\begin{align}
{\cal R}_{\theta} =\exp[i\theta(\hat{a}^\dag\hat{a}-\hat{J}_z)],
\end{align}
which yields
\begin{align}
{\cal R}_{\theta}^\dag(\hat{a},\hat{J}_-
,\hat{J}_+){\cal R}_{\theta} =(\hat{a} e^{-i\theta},\hat{J}_-e^{i\theta},\hat{J}_+e^{-i\theta}).
\end{align}
The occurrence of $U(1)$ symmetry breaking is due to superradiant phase transition.

Superficially, the anti-TCM Hamiltonian (\ref{smantiJC}) is independent of $\Omega_m$ since the rf is not directly coupled to the cavity mode. However, we find that the phase boundary of the superradiant phase transition is affected by the nonzero $\Omega_m$, as shown in the Fig.~\ref{orders} phase diagram in the main text. This result can be understood as follows.

For the vacuum state of cavity $\alpha=0$, the single-particle Hamiltonian of atom-light interaction is reduced to
\begin{eqnarray}
{\cal M} =
\hbar\left(\!
\begin{array}{cc}
U_y\cos^{2}(k_Ly)+ {\delta}/2 & \Omega_m   \\
\Omega_m & U_y\cos^{2}(k_Ly)-{\delta}/2
\end{array}
\!\right)\!,\nonumber \\ \label{smsingleh}
\end{eqnarray}
By diagonalizing the above Hamiltonian, the eigenvalue of the lower branch for the ground state is $\lambda_-=U_y\cos^{2}(k_Ly)-\sqrt{4\Omega_m^2 +\delta^2}/2$, corresponding {\color{blue}to} the eigenstate
\begin{align}
|\chi({\bf r}) = \cos\vartheta |\uparrow\rangle +\sin\vartheta|\downarrow\rangle,
\end{align}
with the mixing angle satisfying the condition $\cot\vartheta =
{\Omega_R}/({\lambda_- - \delta/2}-U_y\cos^{2}(k_Ly))$. Therefore, the populated atom number for the spin-$|\downarrow \rangle$ component in the ground state satisfies
\begin{align}
 N_\uparrow = (\cos\vartheta)^2N_a= \frac{ N_a(\sqrt{\Omega_m^2+\delta^2/4}-\delta/2)}{\sqrt{4\Omega_m^2+\delta^2}},
\end{align}
with $N_a$ being the total atom number of the condensate.

In the thermodynamic limit with $N_{\uparrow}\rightarrow \infty$, the Holstein-Primakoff approach can be introduced by considering the following transformation
\begin{align}
\hat{J}_{-} &= \sqrt{N_{\uparrow} -
\hat{b}^\dag\hat{b}}~\hat{b}, \nonumber \\
\hat{J}_{+} &= \hat{b}^\dag\sqrt{N_{\uparrow} -
\hat{b}^\dag\hat{b}}, \nonumber \\
\hat{J}_{z} &= {\hat b}^\dag{\hat b} -\frac{N_{\uparrow}}{2}, \label{smHP}
\end{align}%
where the bosonic operators $\hat{b}$ and $\hat{b}^\dag$ satisfy the commutation relation $[\hat{b},\hat{b}^\dag]
=1,~[\hat{b},\hat{b}]
=0$, and $[\hat{b}^\dag,\hat{b}^\dag]
=0$, respectively. In the weak excited approximation, e.g., Hamiltonian (\ref{smantiJC}) can be transformed into
\begin{align}
\hat {\cal H}/\hbar &= \tilde{\Delta}_c\hat{a}^\dag\hat{a} +
\omega_0\hat{b}^\dag\hat{b}+
\varepsilon[\hat{a}\hat{b}  + {\rm H.c.}],\label{SMHam}
\end{align}
where $\varepsilon=\Omega\sqrt{{N_{\uparrow}}/{2}}$ is introduced for shorthand notation.

The bilinear Hamiltonian (\ref{SMHam}) in the bosonic operators can be diagonalized by the introduction of a position-momentum representation. Explicitly, the position and momentum operators for the
bosonic cavity and phonon modes are given by ($\hbar=1$ and $M=1$)
\begin{align}
X&=\sqrt{\frac{1}{2\tilde{\Delta}_c}}(\hat a^\dag + \hat
a),~~P_x=i\sqrt{\frac{\tilde{\Delta}_c}{2}}(\hat a^\dag - \hat
a), \nonumber \\
Y&=\sqrt{\frac{1}{2\omega_0}}(\hat
b^\dag + \hat
b),~~P_y=i\sqrt{\frac{\omega_0}{2}}(\hat
b^\dag - \hat b).
\end{align}
Then the Hamiltonian of Eq.~(\ref{SMHam}) in terms of the above operators can be rewritten as $\hat {\cal H}=\hat {\cal H}_r+\hat {\cal H}_p$,
\begin{align}
\hat {\cal H}_r &= \frac{1}{2}\tilde{\Delta}_c^2X^2 +\frac{1}{2}\omega_0^2Y^2
+\varepsilon\sqrt{\tilde{\Delta}_c\omega_0}XY \nonumber\\
&= \frac{1}{2}\lambda_{r+} \tilde{X}^2+\frac{1}{2}\lambda_{r-}
\tilde{Y}^2,
\end{align}%
\begin{align}
\hat {\cal H}_p &= \frac{P_x^2}{2}
+ \frac{P_y^2}{2} -\varepsilon\frac{P_xP_y}{\sqrt{\tilde{\Delta}_c\omega_0}}=  \frac{1}{2}\lambda_{p+} \tilde{P}_x^2+
\frac{1}{2}\lambda_{p-} \tilde{P}_y^2,
\end{align}%
corresponding to the eigenvalues
\begin{align}
\lambda_{r\pm}&= \frac{1}{2}(\tilde{\Delta}_c^2+\omega_0^2)\pm\frac{1}{2}\sqrt{(\tilde{\Delta}_c^2-\omega_0^2)^2 + 4\varepsilon^2\tilde{\Delta}_c\omega_0},\nonumber\\
\lambda_{p\pm}&=
1 \pm \varepsilon{\sqrt{\frac{1}{\tilde{\Delta}_c\omega_0}}} = 1 \pm \sqrt{\frac{\Omega^2N_{\uparrow}}{2\tilde{\Delta}_c\omega_0}}. \nonumber
\end{align}%

To proceed further, we have the final diagonal form
\begin{align}
\hat{H}
&= \frac{1}{2}\lambda_{r+} \tilde{X}^2 + \frac{1}{2}\lambda_{p+} \tilde{P}_x^2 +\frac{1}{2}\lambda_{r-}
\tilde{Y}^2  +
\frac{1}{2}\lambda_{p-} \tilde{P}_y^2 \nonumber\\
&=\lambda_{p+}\sqrt{\lambda_{r+}/\lambda_{p+}}{\cal A}^\dag {\cal A} +
\lambda_{p-}\sqrt{\lambda_{r-}/\lambda_{p-}}{\cal B}^\dag {\cal B}.
\end{align}
We can see clearly that the system will exhibit the instability occurring in the superradiant quantum phase transition when the condition $\Omega^2N_{\uparrow}>2\tilde{\Delta}_c\omega_0$ with both $\lambda_{r-}<0$ and $\lambda_{p-}<0$ simultaneously. Crucially, the critical threshold for Raman coupling strength is
\begin{align}
\Omega_{\rm cr} =\sqrt{{2\tilde{\Delta}_c\omega_{0}/N_\uparrow}}.
\end{align}
Taking into account the cavity decay $\kappa$, the effective cavity detuning $\tilde{\Delta}_c$ should be replaced by $\omega_{\rm eff} = \sqrt{\tilde{\Delta}_c^2+\kappa^2}$ in the position-momentum representation~\cite{baumann2010dicke}. As a result, the critical threshold for Raman coupling strength in the presence of cavity dissipation reads
\begin{align}
\Omega_{\rm cr} =\sqrt{{2(\tilde{\Delta}_c^2+\kappa^2)^{1/2}\omega_{0}/N_\uparrow}}.\label{smPB}
\end{align}
As analyzed in the main text, the analytical equation (\ref{smPB}) is in strong agreement with the numerical results for superradiant quantum phase transition.

\begin{figure}[ht]
\centering\includegraphics[width=0.98\columnwidth]{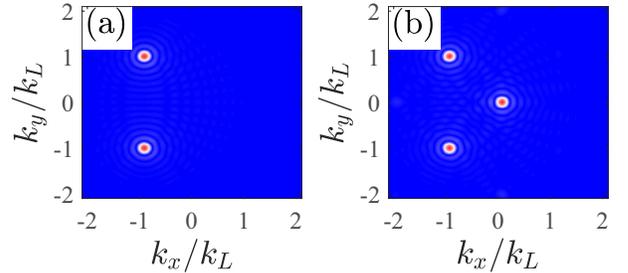}
\protect\caption{The atomic momentum-space distribution ${\cal F}[\psi_{\downarrow}({\bf r})]$ for (a) the self-organized PW phase with $(U_0,\Omega,\Omega_m)=(10,7.5,0)E_L/\hbar$ and (b) the CB supersolid phase  with $(U_0,\Omega,\Omega_m)=(10,7.5,0.5)E_L/\hbar$, respectively. The blue-to-red gradient shading indicates the occupation probability $|\psi_{\downarrow}|$.}
\label{smkspace}
\end{figure}

\begin{figure}[ht]
\centering\includegraphics[width=0.98\columnwidth]{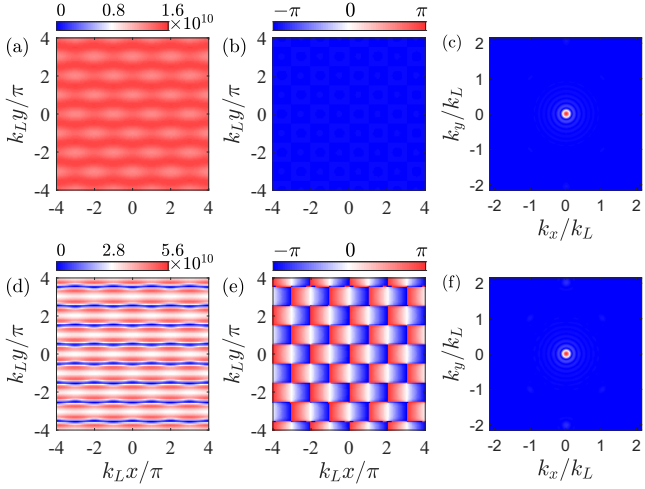}
\protect\caption{The typical densities $\rho_\downarrow$ (column 1), relative phase $\Delta \phi$ (column 2), and atomic momentum-space distribution ${\cal F}[\rho_\downarrow]$ for self-organized NSR phase (upper row) with $(U_0,\Omega,\Omega_m)=(10,6.6,1)E_L/\hbar$ and PW phase (lower row) with $(U_0,\Omega,\Omega_m)=(10,20,1)E_L/\hbar$, respectively.}
\label{smphase}
\end{figure}

Figure~\ref{smkspace} shows the atomic momentum distribution of the condensate wave function of the $|\downarrow\rangle$ state for self-organized plane-wave (PW) and checkerboard (CB) phases in our numerical simulation. As can be seen, the nonzero momenta are dominant in $|k_{x},k_{y}\rangle=|-
\hbar k,\pm \hbar k\rangle$ generated by the dynamical SOC for self-organized superadiant phases, which demonstrates that the single recoil scattering approximation is valid. For the self-ordering CB supersolid phase with $\Omega_m\neq 0$, the additional zero momentum with $|k_{x},k_{y}\rangle=|0,0\rangle$ is expected due to the spatially independent rf field. Remarkably, the self-organization for the CB supersolid phase is manifested by the emergence of the crystal-like periodic density modulation with breaking the continuous translational symmetry. The formation of momentum components at $(\pm \hbar k_L, \pm \hbar k_L)$ for the CB supersolid phase can be ascribed to the interference between the zero momentum $|0,0\rangle$ of the rf field and the $|-\hbar k,\pm \hbar k\rangle$ momentum generated by dynamical SOC [Fig.~\ref{smkspace}(b)].

Figure~\ref{smphase} shows the typical condensate wave functions for the self-organized norma superradiant (NSR) phase (upper row) and PW phase (lower row) in the presence of rf field with $\Omega_m= E_L/\hbar$. It is shown that the condensate density $\rho_\downarrow=|{\psi}_\downarrow|^2$ exhibits a weak stripe of periodic density modulations along the cavity direction for both NSR and PW phases, which are highly consistent with the atomic momentum distribution of atomic condensates, as shown in Figs.~\ref{smphase}(c) and Figs.~\ref{smphase}(f). As for the phase for superradiant states, the relative phase $\Delta \phi=\arg(\psi_\uparrow)-\arg(\psi_\downarrow)$ is homogeneous with $\Delta \phi=-\pi$ for the NSR phase [Fig.~\ref{smphase}(b)] and appears as the staggered $\lambda$-periodic phase modulation along the $x$ axis with $\Delta \phi=-k_L x\pi$ due to the cavity-mediated dynamical SOC [Fig.~\ref{smphase}(d)].

{\section{Effective potential for superradiant phase\label{appB}}}

In order to derive an effective potential for our system, we start by inserting the mean-field ansatz
\begin{align}
\langle\hat{a}\rangle &= \alpha, \nonumber \\
\langle\hat{b}_\downarrow\rangle &=\beta, \nonumber \\
\langle\hat{b}_\uparrow\rangle &= \sqrt{N_{\uparrow}-|\beta|^2},
\end{align}
into the emerged anti-TCM Hamiltonian (\ref{smantiJC}). This results in an effective potential of the cavity-condensate system after adding a constant term:
\begin{align}
V_g(\alpha,\beta) &= \tilde{\Delta}_c|\alpha|^2 + \omega_0|\beta|^2+
\frac{\Omega}{\sqrt{2}}[\alpha\beta\sqrt{N_{\uparrow}-|\beta|^2} + {\rm H.c.}].
\label{potential}
\end{align}
We should emphasize that the cavity field reaches its steady state instantaneously with respect to the atomic field due to $\tilde{\Delta}_c/\omega_0\gg 1$. As a result, we can adiabatically eliminate the cavity field by minimization of $V_g(\alpha,\beta)$ with respect to $\alpha^*$:
\begin{align}
\alpha &=-\Omega\beta^*\sqrt{N_{\uparrow}-|\beta|^2}/\sqrt{2}\tilde{\Delta}_c, \nonumber \\
i\dot{\beta} &=\omega_0\beta+
\frac{\Omega}{\sqrt{2}}\alpha^*\sqrt{N_{\uparrow}-|\beta|^2}.
\end{align}

By substituting $\alpha$ in Eq.~(\ref{potential}), we obtain the expression of the ground-state energy in terms of $V_g(\beta)$:
\begin{align}
V_g(\beta) &=  \omega_0|\beta|^2-
\frac{{\Omega^2}}{2\tilde{\Delta}_c}|\beta|^2(N_{\uparrow}-|\beta|^2) \nonumber \\
&=  \omega_0\left[1 - \frac{{\Omega^2N_{\uparrow}}}{2\omega_0\tilde{\Delta}_c}\right]|\beta|^2+
\frac{{\Omega^2}}{2\tilde{\Delta}_c}|\beta|^4\nonumber \\
&=  \omega_0\left[(1 - \frac{\Omega^2}{\Omega_{cr}^2})|\beta|^2+
\frac{\Omega^2}{N_{\uparrow}\Omega_{\rm cr}^2}|\beta|^4\right]\nonumber \\
&=  \omega_0\left[(1 - \frac{1}{\mu})|\beta|^2+
\frac{1}{\mu N_\uparrow}|\beta|^4\right],
\end{align}
with $\mu=2\tilde{\Delta}_c\omega_0/(N_{\uparrow}\Omega^2)$ for shorthand notation. Clearly, the superradiant quantum phase transition occurs at $\Omega_{\rm cr} =\sqrt{{2\tilde{\Delta}_c\omega_{0}/N_\uparrow}}$ ($\mu=1$) without considering the cavity dissipation.

\begin{figure}[ht]
\centering\includegraphics[width=0.45\columnwidth]{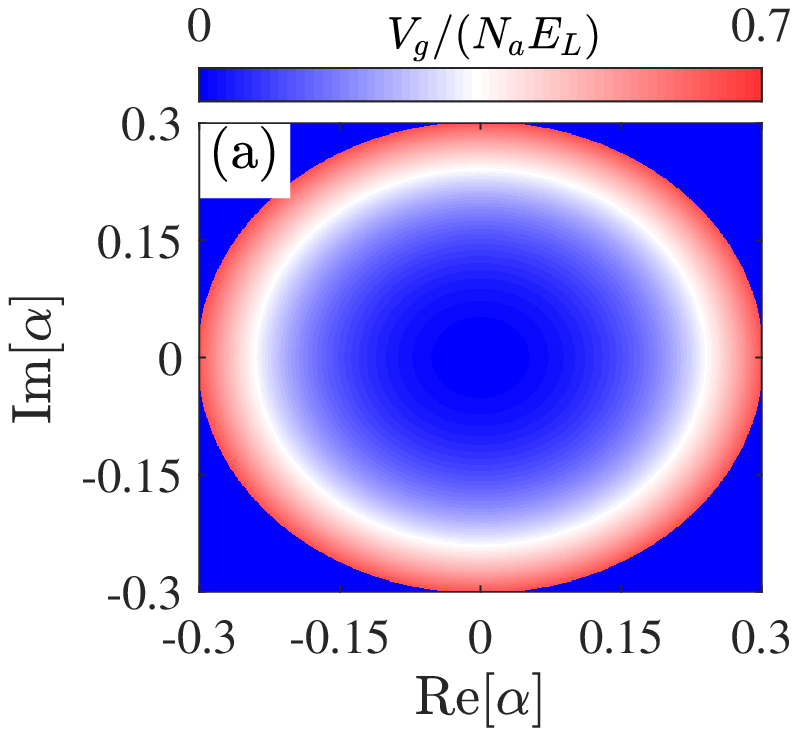}
\centering\includegraphics[width=0.45\columnwidth]{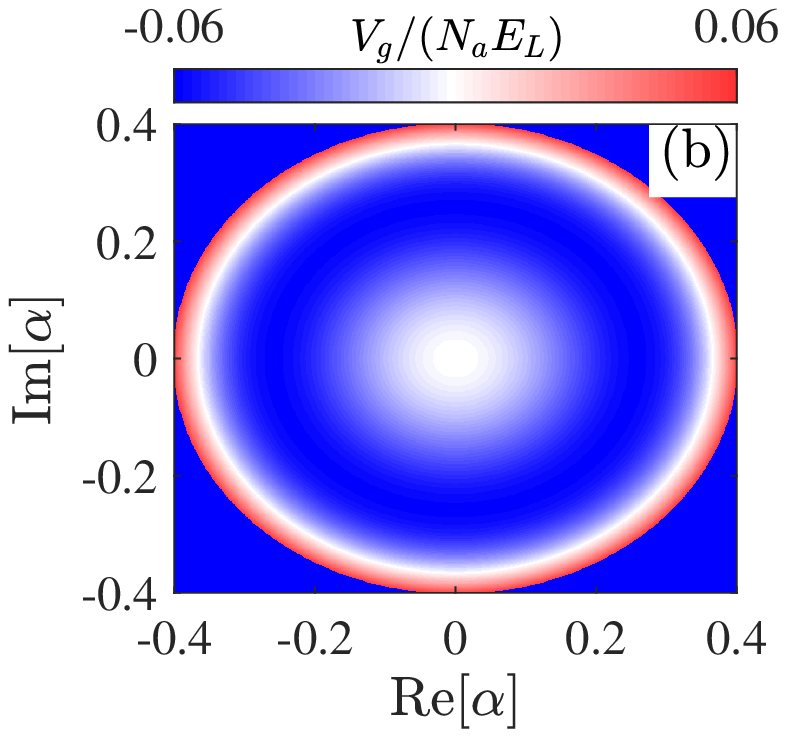}
\protect\caption{The effective potential $V_g$ in (a) normal phase and (b) the superradiant phase as a function of $\alpha$ at $\hbar\Omega/E_L=5$ and $\hbar\Omega/E_L=7.5$, respectively. The blue-to-red color map shows the value of $V_g$. Here the Stark shift is fixed at $\hbar U_0/E_L=10$.}
\label{smpot}
\end{figure}

Figure \ref{smpot} displays the typical effective ground-state potential for the normal phase and superradiant phase. It clear that the ground-state energy emerges with a minimum in the origin for the normal phase and the shape of a Mexican hat with a circular valley of degenerate minima for the superradiant phase, respectively. For the superradiant phase, the emerged sombrero shape with a circular manifold of minima explicitly satisfies
\begin{align}
\alpha &=  -\frac{N_{\uparrow}\Omega}{2\tilde{\Delta}_c}\sqrt{\left(1 - \mu^2\right)/{2}}e^{i\theta}, \nonumber \\
\beta &=  \sqrt{{N_{\uparrow}}\left(1 - \mu \right)/{2}}e^{-i\theta}, \label{smsol}
\end{align}
which spontaneously breaks $U(1)$ symmetry for the superradiant quantum phase transition. Moreover, the potential $V_g(\beta)$ should be experimentally probed along $\beta$ thanks to the expression $\alpha =-\Omega\beta^*\sqrt{N_{\uparrow}-|\beta|^2}/\sqrt{2}\tilde{\Delta}_c$~\cite{leonard2017monitoring}.
\\

\section{Goldstone mode for superradiant phase\label{appC}}

In order to study the collective excitations of the system, we characterize the behavior of the cavity-condensate system by substituting the Holstein-Primakoff transformation of Eq.~(\ref{smHP}) back into the  Hamiltonian (\ref{smantiJC})
\begin{align}
\hat {\cal H}/\hbar &= \tilde{\Delta}_c\hat{a}^\dag\hat{a} +\omega_0\hat{b}^\dag\hat{b} + \frac{\Omega}{\sqrt{2}}[\hat{a} \sqrt{N_{\uparrow} -
\hat{b}^\dag\hat{b}}~\hat{b} + {\rm H.c.}],
\label{JCHP}
\end{align}%
then we displace the bosonic operators $\hat{a}$ and $\hat{b}$ with respect to their mean values in the following ways~\cite{PhysRevE.67.066203}
\begin{align}
\hat{a} &= \alpha + \delta\hat{a}, \nonumber \\
\hat{b} &= \beta + \delta\hat{b},
\end{align}%
where $\delta\hat{a}$ and $\delta\hat{b}$ denotes the photonic and atomic fluctuations of the system around its mean-field values with $\alpha=\langle\hat{a}\rangle$ and $\langle\hat{b}\rangle= \beta$, respectively. For shorthand notation, we {\color{blue}adopt} the notations $\delta\hat{a}\equiv\hat{a}$ and $\delta\hat{b}\equiv\hat{b}$ in the following. Making these displacements, the reduced Hamiltonian for characterizing the excitations of the system is given by
\begin{align}
\hat {\cal H}/\hbar &= \tilde{\Delta}_c(\hat{a}^\dag\hat{a} + \alpha\hat{a}^\dag+\alpha^*\hat{a} +|\alpha|^2)\nonumber \\
&+ \omega_0(\hat{b}^\dag\hat{b} + \beta\hat{b}^\dag+\beta^*\hat{b} +|\beta|^2)  \nonumber \\
&+ \lambda\sqrt{\frac{\zeta}{N_{\uparrow}}}[(\hat{a}+\alpha)\sqrt{\xi}(\hat{b}+ \beta) + {\rm H.c.}],
\label{smHP1}
\end{align}%
with $\lambda= \Omega \sqrt{N_\uparrow/2}$, $\zeta=N_\uparrow-|\beta|^2$, and $\sqrt{\xi}= \sqrt{1- ({\hat{b}^\dag \hat{b} + \beta\hat{b}^\dag+\beta^*\hat{b}})/{\zeta}}$.
In the thermodynamic limit by expanding the square root $\sqrt{\xi}$, the explicit expression reads
\begin{align}
\sqrt{\xi} &= 1- \frac{(\hat{b}^\dag \hat{b} + \beta\hat{b}^\dag+\beta^*\hat{b} )}{2\zeta} - \frac{(\hat{b}^\dag \hat{b} + \beta\hat{b}^\dag+\beta^*\hat{b} )^2}{8\zeta^2}. \nonumber
\end{align}%

By minimizing the ground state energy, the linear terms in the creation (annihilation) operators in Eq.~(\ref{smHP1}) will be zero. As a result, the expanded Hamiltonian (\ref{smHP1}) up to quadratic order in the excitations becomes
\begin{widetext}
\begin{align}
\hat{\cal H}^{(2)}/\hbar &= \tilde{\Delta}_c\hat{a}^\dag\hat{a} +[\omega_0 - \frac{\lambda}{2} \sqrt{\frac{1}{\zeta N_\uparrow}}(\alpha\beta + \alpha^*\beta^*)]\hat{b}^\dag \hat{b} + \frac{\lambda}{2\zeta} \sqrt{\frac{\zeta}{N_\uparrow}}(2\zeta-|\beta|^2)[\hat{a}\hat{b} + \hat{b}^\dag\hat{a}^\dag]\nonumber \\
&-\frac{\lambda|\beta|^2}{2\zeta} \sqrt{\frac{\zeta}{N_\uparrow}}[\hat{b}^\dag \hat{a} e^{-2i\theta} + \hat{a}^\dag\hat{b}e^{2i\theta}]-\frac{\lambda \alpha \beta}{4\zeta^2} \sqrt{\frac{\zeta}{N_\uparrow}}(2\zeta+|\beta|^2)[\hat{b}^\dag e^{-i\theta} + \hat{b}e^{i\theta}]^2,
\end{align}
\end{widetext}
here this quadratic Hamiltonian ${\cal H}^{(2)}$ will determine the excitation spectra of the cavity-condensate system.

To proceed further, we substitute mean-field values $\alpha$ and $\beta$ with the ground-state solution of Eq.~(\ref{smsol}) back into the Hamiltonian ${\cal H}^{(2)}$, and one can obtain an effective Hamiltonian
\begin{align}
\hat{\cal H}^{(2)}/\hbar &= \tilde{\Delta}_c\hat{a}^\dag\hat{a} + \frac{\omega_0}{2\mu}(1+\mu)\hat{b}^\dag \hat{b} + \frac{\Omega'}{4} \frac{1+3\mu}{\sqrt{1+\mu}}[\hat{a}\hat{b} + \hat{b}^\dag\hat{a}^\dag]\nonumber \\
&-\frac{\Omega'}{4} \frac{1-\mu}{\sqrt{1+\mu}}[\hat{b}^\dag \hat{a} e^{-2i\theta} + \hat{a}^\dag\hat{b}e^{2i\theta}] \nonumber \\
&+\frac{\omega_0}{8\mu} \frac{(1-\mu)(3+\mu)}{(1+\mu)}[\hat{b}^\dag e^{-i\theta} + \hat{b}e^{i\theta}]^2.
\label{flua}
\end{align}%
with $\Omega'= \Omega \sqrt{N_\uparrow}$. After the gauge transformations $\hat{b}\rightarrow \hat{b} e^{-i\theta}$ and $\hat{a}\rightarrow \hat{a} e^{i\theta}$, the Bogoliubov Hamiltonian of Eq.~(\ref{flua}) reads
\begin{align}
\hat{\cal H}^{(2)}/\hbar &= \tilde{\Delta}_c\hat{a}^\dag\hat{a} + \frac{\omega_0}{2\mu}(1+\mu)\hat{b}^\dag \hat{b} + \frac{\Omega'}{4} \frac{1+3\mu}{\sqrt{1+\mu}}[\hat{a}\hat{b} + \hat{b}^\dag\hat{a}^\dag]\nonumber \\
&-\frac{\Omega'}{4} \frac{1-\mu}{\sqrt{1+\mu}}[\hat{b}^\dag \hat{a} + \hat{a}^\dag\hat{b}]\nonumber \\
&+\frac{\omega_0}{8\mu} \frac{(1-\mu)(3+\mu)}{(1+\mu)}[\hat{b}^\dag + \hat{b}]^2.
\label{flua1}
\end{align}%
Taking into account the cavity decay $\kappa$, the Hamiltonian of Eq.~(\ref{flua1}) takes the form
\begin{align}
\hat{\cal H}^{(2)}/\hbar &= {\omega}_1\hat{a}^\dag\hat{a} + {\omega}_2\hat{b}^\dag \hat{b} + {\Omega}_1[\hat{a}\hat{b} + \hat{b}^\dag\hat{a}^\dag]\nonumber \\
&+{\Omega}_2[\hat{b}^\dag \hat{a} + \hat{a}^\dag\hat{b}]+{\omega}_3[\hat{b}^\dag + \hat{b}]^2,
\end{align}%
where
\begin{align}
{\omega}_1 &= \tilde{\Delta}_c-i\kappa, \nonumber \\
{\omega}_2 &= \frac{\omega_0}{2\mu}(1+\mu), \nonumber \\
{\omega}_3 &= \frac{\omega_0}{8\mu} \frac{(1-\mu)(3+\mu)}{(1+\mu)}, \nonumber \\
{\Omega}_1 &=  \frac{\Omega'}{4} \frac{1+3\mu}{\sqrt{1+\mu}}\nonumber \\
{\Omega}_2 &=  -\frac{\Omega'}{4} \frac{1-\mu}{\sqrt{1+\mu}},\nonumber
\end{align}
are introduced for shorthand notation. As a result, the Heisenberg equations of motion for the quantum fluctuations in the photonic and atomic field operators can be obtained:
\begin{align}
i\frac{\partial}{\partial t}\hat{a} &={\omega_1}\hat{a} +  \Omega_1\hat{b}^\dag +  \Omega_2\hat{b}, \nonumber \\
i\frac{\partial}{\partial t}\hat{a}^\dag &=-{\omega_1^*}\hat{a}^\dag -\Omega_1\hat{b}-  \Omega_2\hat{b}^\dag, \nonumber \\
i\frac{\partial}{\partial t}\hat{b} &=({\omega_2}+2\omega_3)\hat{b} + \Omega_1\hat{a}^\dag +  \Omega_2\hat{a} + \tilde{\omega}_3\hat{b}^\dag, \nonumber \\
i\frac{\partial}{\partial t}\hat{b}^\dag &=-({\omega_2}+2\omega_3)\hat{b}^\dag - \Omega_1\hat{a}-  \Omega_2\hat{a}^\dag -\tilde{\omega}_3\hat{b}.
\end{align}%

\begin{figure}[ht]
\includegraphics[width=0.98\columnwidth]{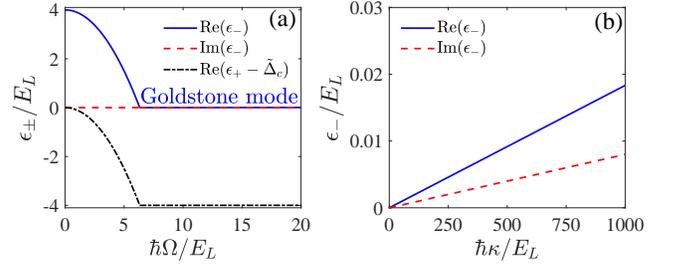}
\caption{(a) The Raman coupling $\Omega$ dependence of collective excitations $\epsilon_{\pm}$ with $\hbar\kappa/E_L=100$. (b) The lower branch of collective excitation $\rm{Re}(\epsilon)_{-}$ and $\rm{Im}(\epsilon_{-})$ as a function of the cavity decay rate $\kappa$ with $\hbar \Omega/E_L=10$. The other parameters are $\hbar U_0/E_L=10$, $\tilde{\Delta}_c=U_0N_a/2=5\times 10^{5}E_L/\hbar$, and $\hbar \omega_0/E_L=4$.}
\label{excitation}
\end{figure}

Furthermore, we recast these equations in the form of the Hopfield-Bogoliubov matrix,
\begin{align}
\left(\begin{array}{cccc}
\omega_1 & \Omega_2 & 0 & \Omega_1 \\
\Omega_2 & \omega_2 + 2\omega_3 & \Omega_1 & \tilde{\omega}_3 \\
0 & -\Omega_1 & -\omega_1^* & -\Omega_2 \\
-\Omega_1 & -\tilde{\omega}_3  & -\Omega_2 & -\omega_2 -2\omega_3
\end{array}\right)\left(\begin{array}{c}
\hat{a} \\
\hat{b} \\
\hat{a}^\dag \\
\hat{b}^\dag
\end{array}\right)=\epsilon \left(\begin{array}{c}
\hat{a} \\
\hat{b} \\
\hat{a}^\dag \\
\hat{b}^\dag
\end{array}\right).
\end{align}
Thus the collective excitation spectra of our model can be conveniently calculated by numerically diagonalizing the Hopfield-Bogoliubov matrix $\hat{\cal H}^{(2)}$.

As to the normal phase with $\alpha=\beta=0$, the bilinear Hamiltonian is given by
\begin{align}
{\cal H}^{(2)}_N &= \tilde{\Delta}_c\hat{a}^\dag\hat{a} +\omega_0\hat{b}^\dag\hat{b} + \lambda[\hat{a}\hat{b} + \hat{a}^\dag\hat{b}^\dag],
\label{normal}
\end{align}%
Then the Heisenberg equations of motion of the quantum fluctuations in the photonic and atomic field operators are given by
\begin{align}
i\frac{\partial}{\partial t}\hat{a} &={\omega_1}\hat{a} +  \lambda\hat{b}^\dag,  \nonumber \\
i\frac{\partial}{\partial t}\hat{a}^\dag &=-{\omega_1^*}\hat{a}^\dag -\lambda\hat{b},\nonumber \\
i\frac{\partial}{\partial t}\hat{b} &={\omega_0}\hat{b} +  \lambda\hat{a}^\dag, \nonumber \\
i\frac{\partial}{\partial t}\hat{b}^\dag &=-{\omega_0}\hat{b}^\dag -  \lambda\hat{a}.
\end{align}%
Analogously, we can recast these equations in the forms of the Hopfield-Bogoliubov matrix,
\begin{align}
\left(\begin{array}{cccc}
\omega_1 & 0 & 0 & \lambda \\
0 & \omega_0 & \lambda & 0 \\
0 & -\lambda & -\omega_1^* & 0 \\
-\lambda & 0  & 0 & -\omega_0
\end{array}\right)\left(\begin{array}{c}
\hat{a} \\
\hat{b} \\
\hat{a}^\dag \\
\hat{b}^\dag
\end{array}\right)=\epsilon \left(\begin{array}{c}
\hat{a} \\
\hat{b} \\
\hat{a}^\dag \\
\hat{b}^\dag
\end{array}\right),
\end{align}%
corresponding to the eigenvalues
\begin{align}
\epsilon_{+}/\hbar&=(\omega_0-\omega_1^* + \sqrt{(\omega_1+\omega_0)^2 - 4\lambda^2})/2, \nonumber\\
\epsilon_{-}/\hbar&=(\omega_1-\omega_0 + \sqrt{(\omega_1+\omega_0)^2 - 4\lambda^2})/2, \nonumber\\
\epsilon_{-}'/\hbar&=(\omega_0-\omega_1^* - \sqrt{(\omega_1+\omega_0)^2 - 4\lambda^2})/2, \nonumber\\
\epsilon_{+}'/\hbar&=(\omega_1-\omega_0 - \sqrt{(\omega_1+\omega_0)^2 - 4\lambda^2})/2.
\end{align}%

We should note that the collective excitations of the system always have two positive and two negative eigenvalues due to the commutation relations of the creation and annihilation operators~\cite{PhysRevLett.112.173601}. The energies of excitations correspond to the non-negative ones with $\epsilon_{+}$ ($\epsilon_{-}$) denoting the higher (lower) branch of collective excitations.

Figure~\ref{excitation}(a) shows the typical collective excitations of the cavity-condensate system as a function of Raman coupling $\Omega$. We find that the energy gap between the higher and lower branches satisfies $|\epsilon_{+}-\epsilon_{-}|\approx \tilde{\Delta}_c=5\times 10^5 E_L/\hbar\gg \omega_0$, which indicates the higher branch $\epsilon_{+}$ is completely decoupled from the ground state of the lower branch $\epsilon_{-}$. As can be seen, the low-energy excitation in our model hosts a gapless Goldstone mode when the Raman field $\Omega$ is above the threshold of the superradiant quantum phase transition, corresponding to the spontaneously broken continuous $U(1)$ symmetry. In particular, the imaging part of the Goldstone mode is roughly zero (the red dashed line). Furthermore, we show that this zero-energy mode is roughly undamped even with a nonzero cavity dissipation when $\kappa/\tilde{\Delta}_c\ll 1$, as displayed in Fig.~\ref{excitation}(b).


%

\end{document}